\documentstyle[aps,prd,12pt]{revtex}
\tightenlines
\newcommand{\beq}{\begin{equation}}
\newcommand{\eeq}{\end{equation}}
\def\beqa{\begin{eqnarray}}
\def\eeqa{\end{eqnarray}}

\input epsf

\begin{document}

\title{A prescription for probabilities in eternal inflation}

\author{Jaume Garriga\( ^{1,2} \) and Alexander Vilenkin\( ^{1} \)}

\address{\( ^{1} \)~Institute of Cosmology, Department of Physics and
Astronomy, Tufts University, Medford, Massachusetts 02155}

\address{\( ^{2} \)~IFAE, Departament de F{\'\i}sica, Universitat
Aut{\`o}noma de Barcelona,\\
08193 Bellaterra \( ( \)Barcelona\( ) \), Spain}

\maketitle
\thispagestyle{empty}

\vspace{2mm}

\begin{abstract}
Some of the parameters we call {}``constants of Nature{}'' may in
fact be variables related to the local values of some dynamical
fields. During inflation, these variables are randomized by
quantum fluctuations.
In cases when the variable in question (call it $\chi$) takes values
in a continuous range,
all thermalized regions in the universe are statistically
equivalent, and a
gauge invariant procedure for calculating the probability
distribution for $\chi$ is known. This is the
so-called {}``spherical cutoff method{}''. In order to find the
probability distribution for $\chi$ it suffices
to consider a large spherical patch in a single thermalized
region. Here, we generalize this method to the case when
the range of $\chi$ is discontinuous and there are several different
types of thermalized region.  We first formulate a set of requirements
that any such generalization should satisfy, and then introduce a
prescription that meets all the requirements.  We finally apply this
prescription to calculate the relative probability for different
bubble universes in the open inflation scenario.
\end{abstract}

\section{Introduction}

The parameters we call {}``constants of Nature{}'' may in fact be variables
related to the local values of certain dynamical fields. For example, what
we perceive as a cosmological constant could be a potential $U(\chi )$ of
some slowly varying field $\chi (x)$. If this potential is very flat, so
that the evolution of $\chi $ is much slower than the Hubble expansion,
then observations will not distinguish between $U(\chi )$ and a true
cosmological constant. Observers in different parts of the universe could
then measure different values of $U(\chi )$.

Spatial variation of the fields $\chi _{a}$ associated with the
{}``constants{}'' can naturally arise in the framework of inflationary
cosmology \cite{1}. The dynamics of light scalar fields during inflation
are strongly influenced by quantum fluctuations, so different regions of
the universe thermalize with different values of $\chi _{a}$. An important
question is whether or not we can predict the values of the
{}``constants{}'' we are most likely to observe. In more general terms, we
are interested in determining the probability distribution
${\mathcal{P}}(\chi )$ for us to measure certain values of $\chi _{a}$. The
answer to this question must involve anthropic considerations to some
extent. The laws of physics may be sufficient to determine the range and
even the spacetime distribution of the variables $\chi _{a}$. However, some
values of $\chi _{a}$ which are physically allowed may be incompatible with
the very existence of observers, and in this case they will never be
measured. The relevant question is then how to assign a weight to this
selection effect.

The inflationary scenario implies a very large universe inhabited by
numerous civilizations that will measure different values of $\chi _{a}$.
We can define the probability ${\mathcal{P}}(\chi )d\chi _{1}...d\chi _{k}$
for $\chi _{a}$ to be in the intervals $d\chi _{a}$ as being proportional
to the number of civilizations which will measure $\chi _{a}$ in that
interval \cite{7}. This includes all present, past and future
civilizations; in other words, it is the number of civilizations throughout
the entire spacetime, rather than at a particular moment of time. Assuming
that we are a typical civilization, we can expect to observe $\chi _{a}$
near the maximum of ${\mathcal{P}}(\chi )$ \cite{8}. The assumption of
being typical has been called the {}``principle of mediocrity{}'' in
Ref.\cite{7}.

An immediate objection to this approach is that we are ignorant
about the origin of life, let alone intelligence, and therefore
the number of civilizations cannot be calculated. But even if
this were true, the approach can still be used to find the
probability distribution for parameters which do not affect the
physical processes involved in the evolution of life. The
cosmological constant $\Lambda $, the density parameter $\Omega $
and the amplitude of density fluctuations $Q$ are examples of
such parameters. Assuming that our fields $\chi _{a}$ belong to
this category, the probability for a civilization to evolve on a
suitable planet is then independent of $\chi _{a}$, and instead
of the number of civilizations we can use the number of habitable
planets or, as a rough approximation, the number of galaxies.
Thus, we can write
\begin{equation}
\label{1} {\mathcal{P}}(\chi
)d^{k}\chi \propto d{\mathcal{N}},
\end{equation}
where $d{\mathcal{N}}$ is the number of galaxies that are going to be
formed in regions where $\chi _{a}$ take values in the intervals $d\chi
_{a}$.

The probability distribution (\ref{1}) based on plain galaxy counting is
interesting in its own right, since it gives a quantitative
characterization of the large scale properties of the universe. Thus, the
general rules for calculating (\ref{1}) are worth investigating quite
independently from anthropic considerations. These considerations can
always be included a posteriori, as an additional factor giving the number
of civilizations per galaxy.

The number of galaxies $d{\mathcal{N}}(\chi )$ in Eq.~(\ref{1}) is
proportional to the volume of the comoving regions where $\chi _{a}$ take
specified values and to the density of galaxies in those regions. The
volumes and the densities can be evaluated at any time, as long as we
include both galaxies that were formed in the past and those that are going
to be formed in the future. It is convenient to evaluate the volumes and
the densities at the time when inflation ends and vacuum energy
thermalizes, that is, on the thermalization surface $\Sigma _{*}$. Then we
can write
\begin{equation} \label{1a}
{\mathcal{P}}(\chi )\propto \nu (\chi ){\mathcal{P}}_{*}(\chi ).
\end{equation}
Here, ${\mathcal{P}}_{*}(\chi )d^{k}\chi $ is proportional to the volume
of thermalized regions where $\chi _{a}$ take values in the intervals
$d\chi _{a}$, and $\nu (\chi )$ is the number of galaxies that form per
unit thermalized volume with cosmological parameters specified by the
values of $\chi _{a}$. The calculation of $\nu (\chi )$ is a standard
astrophysical problem which is completely unrelated to the calculation of
the volume factor ${\mathcal{P}}_{*}(\chi )$, and which does not pose
difficulties of principle.

\begin{figure}[t]
\centering
\hspace*{-4mm}
\leavevmode\epsfysize=10 cm \epsfbox{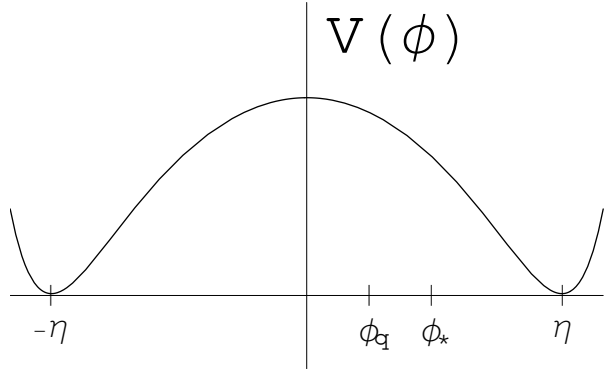}\\[3mm]
\caption[fig1]{\label{fig1} Symmetric double well inflaton potential}
\end{figure}

The meaning of Eq.~(\ref{1}) is unambiguous in models where the total
number of galaxies in the universe is finite. Otherwise, one has to
introduce some cutoff and define the ratio of probabilities for the
intervals $d^{n}\chi ^{(1)}$ and $d^{n}\chi ^{(2)}$ as the ratio of the
galaxy numbers $d{\mathcal{N}}^{(1)}/d{\mathcal{N}}^{(2)}$ in the limit
when the cutoff is removed. However, this limiting procedure has proved to
be rather non-trivial, and a general method which would apply to all
possible eternally inflating scenarios has not yet been found.

The situation is relatively straightforward in the case of an infinite
universe which is more or less homogeneous on very large scales. One can
evaluate the ratio $d{\mathcal{N}}^{(1)}/d{\mathcal{N}}^{(2)}$ in a large
comoving volume ${\mathcal{V}}$ and then take the limit as
${\mathcal{V}}\rightarrow \infty $. The result is expected to be
independent of the limiting procedure; for example, it should not depend on
the shape of the volume ${\mathcal{V}}$. (It is assumed that the volume
selection is unbiased, that is, that the volume ${\mathcal{V}}$ is not
carved to favor some values of $\chi _{a}$ at the expense of other values.)

However, the situation with an infinite universe which is homogeneous on
very large scales is not generic in the context of inflation. Most
inflationary scenarios predict that inflation is eternal to the future, and
therefore the universe is never completely thermalized \cite{AV83,Linde86}
(for a recent review of eternal inflation see \cite{Guth}). An example
which is particularly relevant to the subject of the present paper is given
by the double well inflaton potential depicted in Fig. 1. The inflaton
$\phi $ can thermalize in two different vacua, labeled by $\eta _{1}$ and
$\eta _{2}$. The spacetime distribution of the field in this model is
depicted in Fig. 2. There are thermalized regions of two types,
characterized by the inflaton vacuum expectation value $\eta _{i}$.
Thermalized regions with $\phi =\eta _{1}$ are disconnected from
thermalized regions with $\phi =\eta _{2}$. Both types of regions are
separated by inflating domain walls \cite{Linde94,AV94}, and so the
universe is never completely thermalized. Each thermalization surface is
infinite, so it will contain an infinite number of galaxies. Moreover,
there are an infinite number of thermalized regions of each type.
Therefore, the implementation of Eq. (\ref{1}) for calculating
probabilities requires the comparison of infinite sets of galaxies which
lie in disconnected regions of the universe. A similar spacetime structure
is obtained if we make the inflaton potential periodic by identifying the
two minima. In this case there are still two different types of
thermalization surfaces, characterized by the two topologically different
paths that one can take from the top of the potential to the thermalized
region. Although the particle physics parameters of both types of
thermalized regions are guaranteed to be the same, other cosmological
parameters such as the spectrum of density perturbations or the spatial
distribution of an effective cosmological constant will in general be
different.

\begin{figure}[t]
\centering
\hspace*{-4mm}
\leavevmode\epsfysize=10 cm \epsfbox{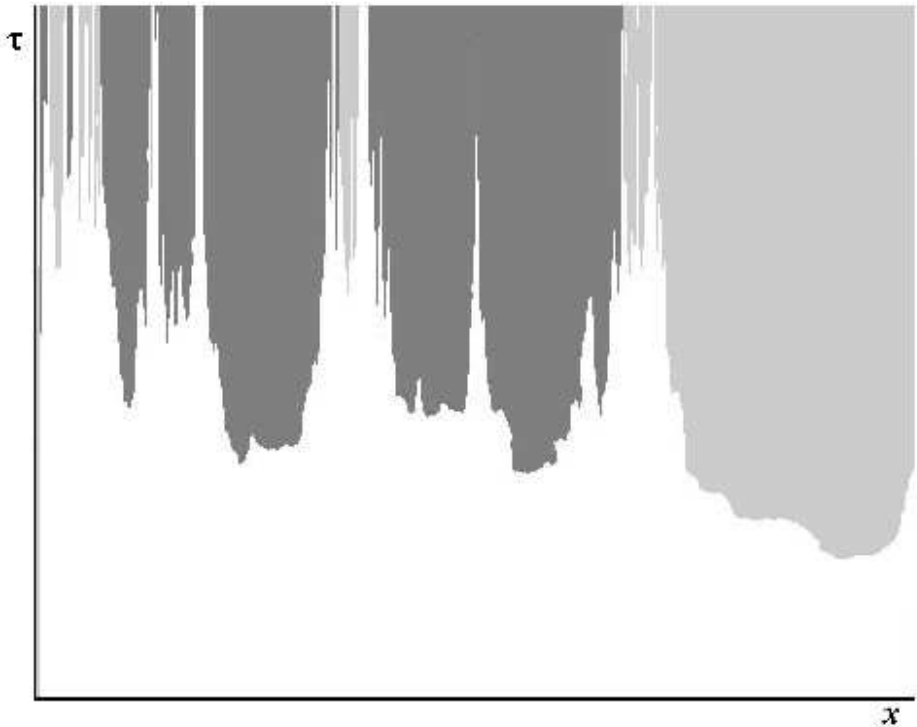}\\[3mm]
\caption[fig2]{\label{fig2} A numerical simulation of the spacetime
structure of an inflating universe \cite{VVW}. The simulation
corresponds to a
double-well inflaton potential, with two degenerate minima where the
inflaton takes the values $\pm \eta$. Inflating regions
are white, while thermalized regions with inflaton values equal to
$+\eta$ and $-\eta$ are shown with different shades of grey .}
\end{figure}

In any model of eternal inflation, the volumes of both inflating and
thermalized regions grow exponentially with time and the number of galaxies
grows without bound, even in a region of a finite comoving size. One can
try to deal with this problem by introducing a time cutoff and including
only regions that thermalized prior to some moment of time $t_{c}$, with
the limit $t_{c}\rightarrow \infty $ at the end. One finds, however, that
the resulting probability distributions are extremely sensitive to the
choice of the time coordinate $t$ \cite{LLM}. Coordinates in General
Relativity are arbitrary labels, and such gauge dependence of the results
casts doubt on any conclusion reached using this approach.

A resolution of the gauge dependence problem was proposed in
Ref.\cite{AV98} and subsequently developed in \cite{VVW}. The proposed
method can be summarized as follows. Let us first assume that inflating and
thermalized regions of spacetime are separated by a single thermalization
surface $\Sigma _{*}$. The problem with the constant-time cutoff procedures
is that they cut the surface $\Sigma _{*}$ in a biased way, favoring
certain values of $\chi $ and disfavoring other values. We thus need a
portion of $\Sigma _{*}$ selected without bias. The simplest strategy is to
use a {}``spherical{}'' cutoff. Choose an arbitrary point $P$ on $\Sigma
_{*}$. Define a sphere of radius $R$ to include all points $Q$ whose
distance from $P$ along $\Sigma _{*}$ is $d(Q,P)\leq R$. We can use
Eq.~(\ref{1}) to evaluate the probability distribution ${\mathcal{P}}(\chi
)$ in a spherical volume of radius $R_{c}$ and then let $R_{c}\rightarrow
\infty $. If the fields $\chi _{a}$ vary in a finite range, they will run
through all of their values many times in a spherical volume of
sufficiently large radius. We expect, therefore, that the distribution
${\mathcal{P}}(\chi )$ will rapidly converge as the cutoff radius $R_{c}$
is increased. We expect also that the resulting distribution will be
independent of the choice of point $P$ which serves as the center of the
sphere. The same procedure can be used for fields with an infinite range of
variation, provided that the probability distributions for $\chi _{a}$ are
concentrated within a finite range, with a negligible probability of
finding $\chi _{a}$ very far away from that range.

Suppose now that there is an infinite number of disconnected thermalization
surfaces, as it happens generically in eternal inflation. Further, we
assume that the variables $\chi _{a}$ of interest are such that their whole
range of values is allowed to occur in a single thermalized region (this is
the case, for instance, for the slowly varying field $\chi $ which plays
the role of a cosmological constant), and that, unlike the case of the
double well potential in Fig.~1, there is only one type of thermalized
region. We can then pick an arbitrary connected component of $\Sigma _{*}$
and apply the spherical cutoff prescription described above. Since the
inflationary dynamics of the fields $\chi _{a}$ have a stochastic nature,
the distributions of $\chi _{a}$ on different connected components of
$\Sigma _{*}$ should be statistically equivalent, and the resulting
probability distribution ${\mathcal{P}}(\chi )$ should be the same for all
components. This has been verified both analytically and numerically in
\cite{VVW}.

The main shortcoming of the spherical cutoff prescription is that as it
stands it cannot be applied to models where the inflaton potential has a
discrete set of minima, as in the example shown in Fig.~1. More precisely,
the problem arises when the minima are separated by inflating domain walls
\cite{Linde94,AV94}. In this case, we can introduce a discrete variable $n$
labeling different minima. Each connected component of the thermalization
surface $\Sigma _{*}$ will be characterized by a single value of $n$
(unless the different minima can be separated by non-inflating domain
walls) and it is clear that the probability distribution for $n$ cannot be
determined by studying one such component.

The purpose of this paper is to propose a generalization of the spherical
cut-off prescription that would be applicable in the general case. We begin
in Section II by formulating the requirements that we believe any such
proposal should satisfy. We require that it should be gauge-independent and
should reduce to the spherical cutoff prescription in the absence of
discrete variables. Moreover, we consider a class of asymmetric double-well
potentials for which the probabilities can be calculated in a
well-motivated way. We then require that the general prescription should
give the same result for this class of potentials. In Section III we
propose a prescription that satisfies all of the above requirements.
We use this prescription in Section IV to calculate
probabilities for bubble universes in open inflation scenario.
Our conclusions are summarised and discussed in Section V.

\section{Requirements}

Suppose that the inflaton potential has $N$ minima labeled by
$i=1,2,...,N$, so that there are $N$ different types of thermalized
regions. Suppose also that there is a set of scalar fields $\chi _{a}$
which take a continuous range of values in all types of thermalized
regions. Our goal is to calculate the probability distribution
\begin{equation} \label{Pn}
{\mathcal{P}}_{i}(\chi )=P_{i}{\hat{\mathcal{P}}}_{i}(\chi ).
\end{equation}
Here, ${\hat{\mathcal{P}}}_{i}(\chi )$ is the normalized distribution
for $\chi $ within $n$-th type of thermalized region,
\begin{equation}
\label{norm}
\int d^{k}\chi {\hat{\mathcal{P}}}_{i}(\chi )=1,
\end{equation}
and $P_{i}$ is the probability for an observer to find herself in a
thermalized region of type $i$.

We begin with the obvious requirements that the probabilities (\ref{Pn})
should be gauge-invariant and should satisfy the normalization condition
\begin{equation} \label{norm'}
\sum _{i}P_{i}=1.
\end{equation}

Next, we require that for $N=1$ the prescription should reduce to the
spherical cutoff prescription. This implies that the spherical cutoff
should be used for the calculation of the individual distributions
${\hat{\mathcal{P}}}_{i}(\chi )$ within each type of thermalized region.
What remains to be determined are the relative probabilities of different
types of regions, $P_{i}$.

\begin{figure}[t]
\centering
\hspace*{-4mm}
\leavevmode\epsfysize=10 cm \epsfbox{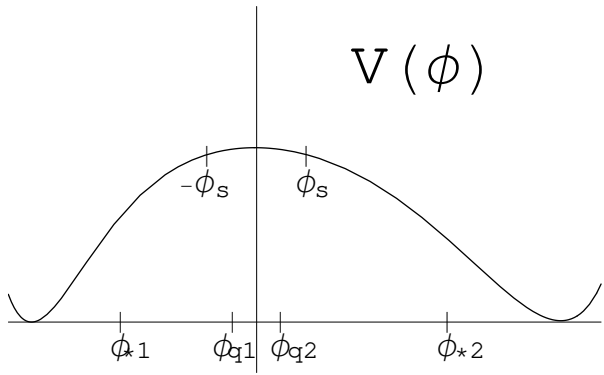}\\[3mm]
\caption[fig3]{\label{fig3} Asymmetric double well potential
which is symmetric near the maximum in
the range $-\phi_s < \phi < \phi_s$.}
\end{figure}

Finally, we introduce a class of inflaton potentials for which we believe
there is a well-motivated answer for the probabilities. Consider first a
symmetric double-well potential, $V(\phi )=V(-\phi )$, with a maximum at
$\phi =0$ and two minima at $\phi _{1,2}=\pm \eta $, as in Fig.~1.
Clearly, in this case
the symmetry of the problem dictates that $P_{1}=P_{2}=0.5$.

Next, we consider an asymmetric double-well, which however \textit{is}
symmetric in some range of $\phi $ near the maximum, $|\phi |<\phi _{s}$
(see Fig.~3). Quantum fluctuations of $\phi $ dominate the dynamics in the
range $|\phi |\lesssim \phi _{q}$, while for $\phi \gg \phi _{q}$
the evolution is essentially deterministic. We shall assume that $\phi
_{s}$ is in the deterministic slow-roll regime, $\phi _{s}\gg \phi _{q}$.
Let us consider constant-$\phi $ surfaces $\phi =\pm \phi _{s}$. These
are infinite spacelike surfaces which have $|\phi |<\phi _{s}$ everywhere
in their past. Since the potential $V(\phi )$ is symmetric in this range
of $\phi $, all these surfaces are statistically equivalent. The symmetry
of the problem suggests that the probabilities $P_{1,2}$ can be calculated
by sampling comoving spheres of equal radius on the two types of surfaces.
The ratio of the probabilities will then be
\begin{equation}
P_{1}/P_{2}=N_{1}/N_{2},
\end{equation}
where $N_{1},N_{2}$ are average numbers of galaxies that will form in
large comoving spherical regions which have equal radii at $\phi =\pm \phi
_{s}$. If we assume for simplicity that the two types of regions have
identical physics at thermalization and afterwards, then the difference
between $N_{1}$ and $N_{2}$ can only be due to the different inflationary
expansion factors $Z_{i}$ characterizing the evolution from $\pm \phi _{s}$
to the thermalization points $\phi _{*i}$. We then have $N_{i}\propto
Z_{i}^{3}$ and
\begin{equation} \label{simple}
P_{1}/P_{2}=(Z_{1}/Z_{2})^{3}.
\end{equation}
We shall require that the general prescription for probabilities should
reproduce Eq. (\ref{simple}) in the case of asymmetric double-well
potentials of the type we discussed here.

\section{The proposal}

In the general case, the inflaton potential has no symmetries to guide our
selection of the equal-$\phi $ surfaces on which to calculate
probabilities. Suppose the potential has a maximum, which we choose to be
at $\phi =0$, and two minima with thermalization points at $\phi _{*1}$ and
$\phi _{*2}$. We can then choose some arbitrary values $\phi _{1}$ and
$\phi _{2}$ in the slow roll ranges of $\phi $ adjacent to $\phi _{*1}$ and
$\phi _{*2}$, respectively, and calculate the probabilities by sampling the
surfaces $\phi =\phi _{i}$. Imagine for a moment that the number of
thermalized regions of both types and the number of galaxies in each region
are all finite. Then we could write
\begin{equation} \label{P1P2}
\frac{P_{1}}{P_{2}}=\frac{p_{1}}{p_{2}}\frac{N_{1}}{N_{2}}.
\end{equation}
Here, $p_{i}$ is the probability for a randomly selected thermalized region
to be of type $i$ and $N_{i}$ is the average number of galaxies in
a type-$i$ region. In our case, however, the number of thermalized regions
and the number of galaxies in each region are infinite, so the definitions
of the probabilities $p_{i}$ and of the ratio $N_{1}/N_{2}$ are
problematic.

It has been remarked \cite{GBL95,Guth} that the problem of determining
$p_{i}$ is similar to the problem of calculating the probability
$p_{\textrm{even}}$ that a randomly selected integer is even. If we take a
long stretch of the natural series
\begin{equation}
\label{natural}
1,2,3,...,
\end{equation}
it will have nearly equal quantities of even and odd numbers, suggesting
that $p_{\textrm{even}}=1/2$. However, the series can be reordered as
\begin{equation} 1,2,4,3,6,8,5,...,
\end{equation}
and the same calculation would give $p_{\textrm{even}}=2/3$. It is clear
that by appropriately ordering the series one can obtain any answer for
$p_{\textrm{even}}$ between $0$ and $1$. This seems to suggest that the
probabilities $p_{i}$ are hopelessly ill-defined.

We note, however, that the situation with the natural series is not as bad
as it may seem. The series has a natural ordering in which the nearest
neighbors of each number differ from that number by 1, and we can require
that our sampling procedure should respect this natural {}``topology{}''.
Then we have $p_{\textrm{even}}=1/2$, which is the answer that one
intuitively expects. With an infinite number of thermalized regions of
different types, one could also order the list of regions in a way that
would give any desired result for $p_{1}/p_{2}$. But again one can hope
that this ambiguity can be removed if we require that our sampling
procedure should reflect the spatial distribution of the regions.

The ratio $N_{1}/N_{2}$ can be calculated by counting galaxies in spheres
located in regions of the two types. In the double-well example of Section
II, there is complete symmetry between the surfaces $\phi =\phi _{s}$ and
$\phi =-\phi _{s}$. We expect, therefore, that $p_{1}=p_{2}$ and we choose
the radii of the spheres to be equal at $\phi =\pm \phi _{s}$. However,
it is not
clear what sets the relative size of the spheres in the absence of
symmetry. We shall now describe the method we propose for evaluating
$p_{i}$ and $N_{1}/N_{2}$ in the general case.

We start by noting that there is one thing that thermalized regions of the
two types have in common. In their past they all went through a period of
stochastic inflation, with the inflaton field $\phi $ undergoing a random
walk near the top of the potential. Our idea is to use some markers from
this early era for the calculation of probabilities.

Let us imagine that markers are point objects and that they are produced at
a constant rate per unit spacetime volume in inflating regions where $\phi
$ is at the top of its potential, $V(\phi )\approx V_{max}$. After that,
the markers evolve as comoving test particles and eventually end up in a
thermalized region of one type or the other. We shall define $p_{i}$ as the
fraction of markers that end up in regions of type $i$. Furthermore, the
reference length scale on which the number of galaxies is counted in a
region of type $i$ will be set by the average distace $d_{i}$ between
markers in that type of region. In other words, the galaxies are counted in
spheres of radii $R_{1}$ and $R_{2}$ such that $R_{1}/R_{2}=d_{1}/d_{2}$.
To summarize, we propose that the relative probabilities for the constants
of nature are given by (\ref{Pn}) with
\begin{equation}
\label{proposal}
\frac{P_{1}}{P_{2}}=\frac{p_{1}n_{2}}{p_{2}n_{1}}\frac{\nu _{1}}{\nu _{2}},
\end{equation}
where $n_{i}\propto d_{i}^{-3}$ is the mean number density of markers
in region of type $i$ and $\nu _{i}$, given by \[
\nu _{i}^{-1}=\int \hat{\mathcal{P}}_{i}(\chi )\nu _{i}^{-1}(\chi )d\chi \]
is the mean density of galaxies in that type of region. As mentioned in the
introduction, the calculation of $\nu _{i}$ is a standard astrophysical
problem which we shall not be concerned with in this paper.

A physical counterpart to the production of our idealized markers is
quantum nucleation of black holes in an inflating universe. Black holes are
produced at a rate \cite{GP} $r\propto \exp [-1/8V(\phi )]$ which grows
exponentially with $V(\phi )$, so that by far the highest rate is achieved
at $V(\phi )=V_{max}$. Here, we prefer not to identify markers with black
holes and to think of them as of idealized point objects. This frees us
from concerns such as the contribution to black hole production from
regions which are not at the top of the potential, black hole evaporation,
etc. The reality and observability of markers is not really an issue, since
the comparison of causally disconnected thermalized regions is certainly a
gedanken experiment.

The average comoving distance between the markers in a thermalized region
is manifestly gauge invariant, and so is the ratio $n_{2}/n_{1}$ in Eq.
(\ref{proposal}). The quantities $p_{i}$ are also gauge-independent, and
thus the probabilities $P_{i}$ in (\ref{proposal}) should be
gauge-invariant.

It is also easy to verify that the above proposal gives the expected result
(\ref{simple}) for the probabilities when the inflaton potential is
symmetric in the diffusion range of $\phi $. In this case, the average
distance between markers on the surface $\phi =\phi _{s}$ is the same as
that on $\phi =-\phi _{s}$ (with $\phi _{s}$ defined in Section II). Hence,
the distances between markers at thermalization may differ only due to the
difference in the expansion factors from $\pm \phi _{s}$ to $\phi _{*i}$,
\begin{equation} d_{1}/d_{2}=Z_{1}/Z_{2},
\end{equation}
and Eq.(\ref{simple}) follows immediately.


\begin{figure}[t]
\centering
\hspace*{-4mm}
\leavevmode\epsfysize=10 cm \epsfbox{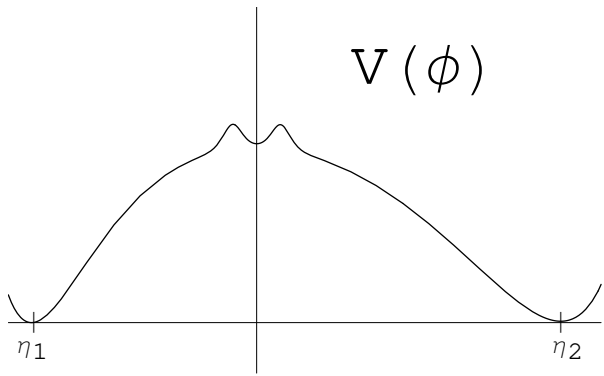}\\[3mm]
\caption[fig4]{\label{fig4}
Asymmetric inflaton potential with a metastable vacuum
at the top, which replaces the quantum diffusion region of
Fig. 3.}
\end{figure}

\section{Open inflation}

As an illustration of the method we shall now calculate probabilities
for a model of {}``open inflation{}'' \cite{Gott,Bucher,Tanaka}. We assume
an inflaton potential of the form shown in Fig.~4. It has a metastable
false vacuum at $\phi =0$ which is separated by potential barriers from two
slow roll regions on the left and on the right. The false vacuum decays
through bubble nucleation, and the inflaton rolls towards the true vacuum
inside the bubbles. Comoving observers inside each bubble would, after
thermalization of the inflaton, see themselves in an open homogeneous
universe. (Hence the name {}``open inflation{}''.) If the bubble nucleation
rate is not too high, bubble collisions are rare and inflation is eternal.
Assuming that the two types of bubbles have identical low-energy physics,
we would like to find the probability for an observer to find herself in
one type of bubble or the other.

In false vacuum regions outside bubbles the metric is de Sitter,
\begin{equation} ds^{2}=dt^{2}-e^{2H_{0}t}d{\mathbf x}^{2}
\end{equation}
where $H_{0}=8\pi GV_{0}/3$ and $V_{0}=V(0)$ is the false vacuum
energy density. The bubble interior has the geometry of an open
Robertson-Walker universe, \begin{equation}
\label{RW}
ds^{2}=d\tau ^{2}-a^{2}(\tau )[d\xi ^{2}+\sinh ^{2}\xi d\Omega ^{2}].
\end{equation}
After nucleation, the bubble wall expands rapidly approaching the speed of
light. If the initial bubble size is much smaller than the de Sitter
horizon $H_{0}^{-1}$, then the wall worldsheet is well approximated by the
future light cone of the center of spacetime symmetry of the bubble. We
shall choose coordinates so that this center is at ${\mathbf x}=t=0$ and
$\tau =0$. Then, as $t\rightarrow \infty $, the bubble wall asymptotically
approaches $|{\mathbf x}|\rightarrow H_{0}^{-1}$. The spacetime geometry of
the bubble is illustrated in Fig.~5.

The relation between the coordinates $(t,{\mathbf x})$ and $(\tau ,\xi )$
can be easily found if we assume, as it is usually done, that (i) the
potential $V(\phi )$ immediately outside the barrier has nearly the same
value $V_{0}$ as in the false vacuum, and (ii) that the gravitational
effect of the bubble wall is negligible. Then, at sufficiently small values
of $\tau $, the metric (\ref{RW}) inside the bubble is close to the open de
Sitter metric, $a(\tau )=H_{0}^{-1}\sinh (H_{0}\tau )$. The coordinates
$(t,{\mathbf x})$ and $(\tau ,\xi )$ are related by the usual
transformation between the flat and open de Sitter charts, which we shall
not reproduce here.

We assume that markers are produced at a constant rate $r$ in the false
vacuum outside the bubbles. The density of markers $n({\mathbf x},t)$
satisfies the equation \begin{equation}
\frac{\partial n}{\partial t}+3H_{0}n=r,
\end{equation}
which has a stationary solution $n_{0}=r/3H_{0}$. This de Sitter-invariant
solution is rapidly approached, regardless of the initial conditions. We
are interested in the average separation between the markers on the
surfaces $\tau =\tau _{*i}$ corresponding to thermalization of the inflaton
in the two types of bubbles. The densities of markers $n_{*i}$ on these
surfaces are also constant, due to the spacetime symmetry of the bubbles
(all points on the surface $\tau ={\textrm{const}}$ are equivalent).

\begin{figure}[t]
\centering
\hspace*{-4mm}
\leavevmode\epsfysize=10 cm \epsfbox{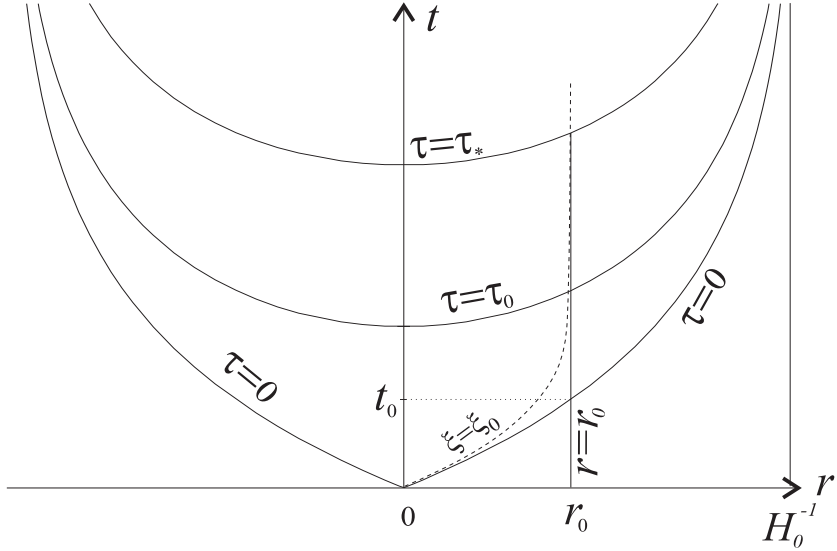}\\[3mm]
\caption[fig5]{\label{fig5}
Spacetime geometry of the bubble in co-moving coordinates. Here,
$r=|{\bf\rm x}|$, and various $\tau=const.$ surfaces of the open
Robertson-Walker universe inside the bubble are shown, as well as
one of the $\xi=const.$ curves.}
\end{figure}

The interior geometries of different types of bubbles are nearly identical
at early times (small $\tau $), while $V(\phi )\approx V_{0}$. Let us
choose a value $\tau _{0}$ in that range and consider surfaces $\tau =\tau
_{0}$. The geometry and the distribution of markers in the past of such
surfaces are the same for the two types of bubbles, and therefore the
average separations of markers on these surfaces should also be the same,
$d_{1}(\tau _{0})=d_{2}(\tau _{0})\equiv d_{0}$. The separation of markers
at thermalization is $d_{*i}=Z_{*i}d_{0}$, where $Z_{*i}$ is the expansion
factor between $\tau _{0}$ and $\tau _{*i}$, and we have \begin{equation}
\label{Nb}
N_{1}/N_{2}=(Z_{*1}/Z_{*2})^{3}.
\end{equation}

To complete the calculation, we have to determine the fraction $p_{i}$
of the markers that end up in type-$i$ bubbles. Let $f_{i}(t)$ be
the fraction of comoving volume occupied by type-$i$ bubbles in a comoving
volume which we assume to be free of bubbles at the initial moment $t=0$,
\begin{equation}
\label{f0}
f_{i}(0)=0.
\end{equation}
The evolution of $f_{i}(t)$ is described by the equations \begin{equation}
\frac{\partial f_{1}}{\partial t}=\frac{4\pi \lambda
_{1}}{3H^{3}}(1-f_{1}-f_{2}),
\end{equation}
\begin{equation}
\frac{\partial f_{2}}{\partial t}=\frac{4\pi \lambda
_{2}}{3H^{3}}(1-f_{1}-f_{2}),
\end{equation}
where $\lambda _{i}$ is the nucleation rate of type-$i$ bubbles.
In these equations we are neglecting {}``secondary{}'' bubble nucleation
which may occur within a comoving distance $H^{-1}$ from any given
{}``primary{}'' nucleation event, before the primary bubble reaches its
asymptotic cooving size $H^{-1}$. Secondary nucleations will affect the
co-moving volume distribution very little, especially if the nucleation
rate per unit volume is much smaller than $H^{4}$. This is a typical
situation since the nucleation rates are usually exponentially suppressed.
The solution of these equations with the initial condition (\ref{f0}) is
\begin{equation} f_{i}(t)=\frac{\lambda _{i}}{\lambda }\left[ 1-\exp \left(
-\frac{4\pi \lambda }{3H^{3}}t\right) \right] ,
\end{equation}
where $\lambda =\lambda _{1}+\lambda _{2}$. The fraction of markers that
end up in bubbles of type $i$ is given by \begin{equation}
\label{pb}
p_{i}=f_{i}(t\rightarrow \infty )=\lambda _{i}/\lambda .
\end{equation}

Now, substituting (\ref{Nb}) and (\ref{pb}) into Eq.(\ref{P1P2}) for the
probabilities we obtain \begin{equation}
\label{pbubble}
\frac{P_{1}}{P_{2}}=\frac{\lambda _{1}}{\lambda _{2}}\left(
\frac{Z_{*1}}{Z_{*2}}\right) ^{3}.
\end{equation}
This agrees with one's intuitive expectation that the probability should be
proportional to the nucleation rate and to the volume expansion factor
inside the bubbles.

\section{Discussion}

In this paper we have suggested a cutoff procedure which allows one to
assign probabilities to different types of thermalized regions in an
eternally inflating universe. The probabilities are calculated with the aid
of {}``markers{}'' -- imaginary pointlike objects which are assumed to be
created at a constant rate in the inflating regions where the inflaton
field is at the top of its potential. The probability for regions of type
$i$ is then \begin{equation} P_{i}\propto p_{i}N_{i},
\end{equation}
where $p_{i}$ is the fraction of markers that end up in type-$i$
regions and $N_{i}$ is the number of galaxies formed in a comoving sphere
of radius equal to the average separation between the markers.

In contrast to some earlier proposals, the new prescription is manifestly
gauge-invariant. It also gives the expected results in cases where we have
well-motivated intuitive expectations for the probabilities.

The method of calculating probabilities presented in this paper has some
similarities to the so-called $\epsilon $-prescription which was proposed
in Ref. \cite{AV95}. Starting with a comoving volume with $\phi $ near the
top of the potential, the numbers of galaxies are calculated in this
prescription by imposing cutoffs at different times in different types of
regions. The cutoff in type-$i$ regions is chosen at the time $t_{ci}$ when
all but a small fraction $\epsilon $ of the comoving volume destined to
thermalize in this type of regions has thermalized. The value of $\epsilon
$ is the same for all types of regions, but the cutoff times are
different. The limit $\epsilon \rightarrow 0$ is taken after calculating
the probabilities.  The $\epsilon$-prescription was applied in
\cite{VW97} to calculate the probabilities in open inflation and gave
the same result (\ref{pbubble}) that we obtained here.

To see the connection between this prescription and the method of the
present paper, imagine that the initial comoving region contains a large
number of markers. If $p_{i}$ is the fraction of markers that are going to
end up in type-$i$ regions, then the cutoffs in $\epsilon $-prescription
are imposed at the times when the numbers of markers ${\mathcal{N}}_{mi}$
in the two types of regions are related by
${\mathcal{N}}_{m1}/{\mathcal{N}}_{m2}=p_{1}/p_{2}$. The number of galaxies
in each type of region is ${\mathcal{N}}_{gi}={\mathcal{N}}_{mi}N_{i}$,
where $N_{i}$ is the number of galaxies per one marker. Hence,
\begin{equation}
\frac{P_{1}}{P_{2}}=\frac{{\mathcal{N}}_{g1}}{{\mathcal{N}}_{g2}}=\frac{p_{1}}{p_{2}}\frac{N_{1}}{N_{2}},
\end{equation}
which has the same form as (\ref{P1P2}).

One difference between the two methods is that markers are continuously
produced in our new approach, while in $\epsilon $-prescription new markers
are not produced even in regions when the inflanton field fluctuates back
to the top of its potential. Another difference is that $\epsilon$
-prescription uses constant-$t$ cutoffs, while the new approach uses
spherical cutoffs. Because of its reliance on a constant-$t$ cutoff, the
results obtained using $\epsilon $-prescription are generally
gauge-dependent, whereas the new method is gauge independent.

The most straightforward way to implement the new prescription is through a
numerical simulation of an eternally inflating spacetime. This method,
however, suffers from severe computational limitations
\cite{VVW}. Alternatively, one can use an approximate analytic method
based on the Fokker-Planck (FP) formalism of stochastic inflation
\cite{AV83,Starob,LLM}.  This method works very well for the
calculation of $p_i$ \cite{GVV}.  However, the results for the density
of markers obtained using this approach are generally not gauge invariant.

The FP formalism can be used to calculate the physical volume
$V_{*i}(t)$ that thermalizes prior to some time $t$ and the number of
markers $N_{*i}(t)$ contained in that volume for different types of regions.
The average distance between the markers can be expressed as
\begin{equation} \label{dt}
d_{i}(t)=\left[ \frac{V_{*i}(t)}{N_{*i}\left( t\right) }\right] ^{1/3}.
\end{equation}
One might expect that in the limit $t\rightarrow \infty $, $d_{i}(t)$
should approach the average separation between markers on the
thermalization surfaces $\Sigma _{*i}$ (calculated in large spherical
volumes). However, this is not generally the case. The thermalized volume
$V_{*i}(t)$ in an eternally inflating universe is dominated by the newly
thermalized regions (shortly before the time $t$), and the density of
markers in these regions is generally correlated with the choice of the
time variable $t$.

The density of markers on the surfaces $\Sigma _{*i}$ can vary greatly
from one area to another. In some places the field $\phi $ takes a more
or less direct route from the top of the potential to $\phi _{*i}$,
resulting in a relatively high density of markers, while in others it takes
a long time fluctuating up and down the potential, so that the markers are
greatly diluted. By including only regions that thermalize prior to time
$t$, we {}``reward{}'' regions that thermalize faster and therefore
introduce a bias favoring higher densities of markers. This qualitative
tendency is present for most choices of the time variable, but
quantitatively the bias will not be the same. Hence, one should not be
surprised that $d_{i}$ calculated from Eq.(\ref{dt}) would depend on the
choice of gauge.

Despite this gauge dependence, the FP method may give
approximately valid results for some class of models.
It has been argued in Refs. \cite{AV95,WV96} that in the case of
$\epsilon$-prescription the gauge dependence is rather weak for a wide class
of potentials.  One can expect the situation to be similar for our new
method when a constant $t$ cutoff is used.  However, more work is needed
to determine what additional requirements the potential should satisfy
for this approximate gauge independence to apply.

We started in Section II of this paper by formulating a set of
requirements which we believe any method for calculating the
probabilities should satisfy.  The specific prescription we introduced
in Section III can be regarded as an ``existence proof'',
demonstrating that a prescription satisfying all the criteria can
indeed be constructed.  It is quite possible, however, that our
prescription is not unique and that more attractive and better
motivated methods can be developed.  With this in mind, we conclude by
indicating some possible shortcomings and limitations of our method.

Admittedly, our prescription includes an element of arbitrariness when we
assume that markers are produced only in regions where the inflaton $\phi $
is at the top of its potential. This {}``$\delta$-function{}'' source
should not be understood literally. The semiclassical picture of eternal
inflation involves smearing over spacetime scales $\sim H^{-1}$ and over
scalar field intervals $\sim H/2\pi $. Hence, marker formation at the top
of the potential is equivalent to marker formation within an interval
$\Delta \phi \sim H/2\pi $ from the top of the potential. However, an
attempt to extend marker formation to a wider interval encounters some
ambiguities. Formally, there is no problem in calculating the density
$n_{i}$ of markers in thermalized regions even if these are produced at
some given rate $R(\phi )d\phi $ per unit proper time and volume in regios
where the field is in the range $d\phi $. However, there is an obvious
arbitrariness in what should be chosen as our {}``smearing function{}''
$R(\phi )$. Also, if markers are produced at different rates at different
values of $H(\phi )$, the calculation of the fraction of markers $p_{i}$
that end up in region $i$ becomes somewhat ill posed. Markers formed at
different values of $\phi $ will have a different probability $p_{i}$ of
ending up in a type-$i$ region and it is not clear how to weigh the
different contributions. One could calculate $p_{i}(\phi )$ and then
average over $\phi $ with some weight, but it is not clear how the weight
function is to be determined.  It thus appears that confining
marker formation to the top of the potential has some advantages and
may be not as arbitrary as it first seems.

It is not clear whether or not our method will give reasonable results
when applied to the most general type of inflationary scenarios, when
the inflaton potential has several local maxima.  Additional
complications arise in models where the minimum of the potential has a
non-vanishing cosmological constant. In such models, regions of true vacuum
in the post-inflationary universe may fluctuate back to the quantum
diffusion range of the inflaton potential and the spacetime structure
is more complicated than the one represented in Fig.~2.

Also, our method is not applicable when the potential is unbounded from
above, in which case eternal inflation runs into the Planck boundary. This
is not particularly worrisome, since after all the inflaton may be a
modulus with a compact range, and the inflaton potential may well be
bounded from above at a scale much lower than the Planck scale.

\section*{acknowledgements}

It is a pleasure to thank Ken Olum, Masafumi Seriu and Serge Winitzki
for useful
discussions. This work was supported by the Templeton Foundation under
grant COS 253. J.G. is partially supported by CICYT, under grant
AEN99-0766. A.V. is partially supported by the National Science Foundation.

\end{document}